\documentclass[3p,times,procedia]{elsarticle}
\usepackage{nupha_ecrc}

\usepackage{graphicx}
\usepackage{amssymb}
\usepackage{dcolumn}
\usepackage{amsmath}
\usepackage{bm}
\usepackage{textcomp}
\usepackage{epstopdf}
\usepackage{color}
\usepackage{ulem}
\usepackage[toc,page,title,titletoc,header]{appendix}
\usepackage[colorlinks,
            linkcolor=blue,
            anchorcolor=blue,
            citecolor=blue
            ]{hyperref}
\usepackage{txfonts}
\usepackage{caption}
\usepackage{multirow}

\volume{00}

\firstpage{1}

\journalname{Nuclear Physics A}

\runauth{}

\jid{nupha}

\jnltitlelogo{Nuclear Physics A}

\usepackage{amssymb}
\usepackage[figuresright]{rotating}
\begin{document}

\begin{frontmatter}

\dochead{}

\title{Multiplicity fluctuations of net protons on the hydrodynamic freeze-out surface}
 %at $\sqrt{s_{NN}}$ = 7.7, 11.5, 19.6,27, 39, 62.4 and 200 GeV}

\author[label1]{Lijia Jiang}
\author[label1]{Pengfei Li}
\author[label1,label2,label3]{Huichao Song}
\address[label1]{Department of Physics and State Key Laboratory of Nuclear Physics and
Technology, Peking University, Beijing 100871, China}
\address[label2]{Collaborative Innovation Center of Quantum Matter, Beijing 100871, China}
\address[label3]{Center for High Energy Physics, Peking University, Beijing 100871, China}


\begin{abstract}
This proceeding briefly summarizes our recent work on calculating the correlated fluctuations of net protons on the hydrodynamic freeze-out surface  near the QCD critical point. For both Poisson and Binomial baselines, our calculations could roughly reproduce the energy
dependent cumulant $C_4$ and $\kappa \sigma^2$ of net protons, but always over-predict $C_2$ and $C_3$ due to the positive contributions from the static critical fluctuations.
\end{abstract}

\end{frontmatter}

\section{Introduction}

One of the fundamental goals of Relativistic Heavy Ion Collisions (RHIC)
is to locate the critical point of the QCD phase diagram~\cite{Aggarwal:2010cw}.
To search the critical point in experiment, the STAR collaboration has systematically measured
the higher cumulants of net-protons in Au+Au collisions at $\sqrt{s_{NN}}$ = 7.7, 11.5, 19.6,
27, 39, 62.4 and 200 GeV~\cite{Aggarwal:2010wy, Adamczyk:2013dal,Luo:2015ewa}. With the maximum transverse momentum increased from
$0.8$ to $2$ GeV, the measured cumulant ratios $\kappa \sigma ^{2}$ of net protons show large
deviations from the Poisson baselines and present an obvious non-monotonic behavior in the most
central Au+Au collisions~\cite{Luo:2015ewa}, which hints the signal of the QCD critical point. To quantitatively
study these experimental data, we need to develop
dynamical models near the QCD critical point. Currently, most of the dynamical models near
the QCD critical point, e.g. chiral fluid dynamics, focus on the dynamical evolution of the bulk matter
~\cite{Paech:2003fe}. In a recent paper~\cite{Jiang}, we introduced a freeze-out scheme for the
dynamical models near the QCD critical point through coupling the decoupled classical particles with the order parameter field.
With a modified distribution function, we calculated the correlated fluctuations of net protons on the hydrodynamic freeze-out
surface in Au+Au collisions at various collision energies. In this proceeding, we will
briefly summarize the main results of that paper.

\section{The formalism and set-ups}
In traditional hydrodynamics, the particles emitted from the freeze-out
surface can be calculated through the Cooper-Frye formula with a classical
distribution function $f(x,p)$~\cite{Song:2007fn}. In the vicinity of the critical point, we assume
the effective mass of the classical particles strongly fluctuates through
interacting with the order parameter field: $\delta m =g\sigma(x)$, which introduces a modified distribution function that
also correlated fluctuates in position space~\cite{Jiang}.  To the linear order of $\sigma(x)$, the modified distribution
function can be expanded as:
\begin{equation}
f=f_{0}+\delta f=f_{0}\left( 1-g\sigma /\left( \gamma T\right) \right) ,
\end{equation}%
where $f_{0}$ is the traditional equilibrium distribution function,
$\delta f$ is the fluctuation deviated from the equilibrium part,
$\gamma =\frac{p^{\mu }u_{\mu }}{m}$ is the covariant Lorentz factor and
the coupling constant $g=\frac{dm}{d\sigma }$. With such expansion, the connected 2-point 3-point and 4-point
correlators $\left\langle \delta f_{1}\delta f_{2}\right\rangle _{c}$,
$\left\langle \delta f_{1}\delta f_{2} \delta f_{3}\right\rangle _{c}$ and
$\left\langle \delta f_{1}\delta f_{2} \delta f_{3} \delta f_{4}\right\rangle _{c}$
are proportional to $\left\langle \sigma _{1}\sigma _{2}\right\rangle _{c}$,
$\left\langle \sigma _{1}\sigma _{2}\sigma_{3}\right\rangle _{c}$ and
$\left\langle \sigma _{1}\sigma _{2}\sigma_{3}\sigma_{4}\right\rangle _{c}$, respectively.
Integrating over the whole freezeout surface for theses correlators gives the explicit forms of the critical
fluctuations for produced hadrons~\cite{Jiang}:
\begin{align}
\left\langle \left( \delta N\right) ^{2}\right\rangle _{c}=& \left( \frac{%
1}{\left( 2\pi \right) ^{3}}\right) ^{2}\prod_{i=1,2}\left( \int \frac{1%
}{E_{i}}d^{3}p_{i}\int_{\Sigma _{i}}p_{i\mu }d\sigma _{i}^{\mu }\right)
\frac{f_{01}f_{02}}{\gamma _{1}\gamma _{2}}\frac{g^{2}}{T^{2}}%
\left\langle \sigma _{1}\sigma _{2}\right\rangle _{c}, \\
\left\langle \left( \delta N\right) ^{3}\right\rangle _{c}=& \left( \frac{%
1}{\left( 2\pi \right) ^{3}}\right) ^{3}\prod\limits_{i=1,2,3}\left(
\int \frac{1}{E_{i}}d^{3}p_{i}\int_{\Sigma _{i}}p_{i\mu }d\sigma _{i}^{\mu
}\right)
\frac{f_{01}f_{02}f_{03}}{\gamma _{1}\gamma _{2}\gamma _{3}}\left(
-1\right) \frac{g^{3}}{T^{3}}\left\langle \sigma _{1}\sigma _{2}\sigma
_{3}\right\rangle _{c}, \\
\left\langle \left( \delta N\right) ^{4}\right\rangle _{c}=& \left( \frac{%
1}{\left( 2\pi \right) ^{3}}\right) ^{4}\prod\limits_{i=1,2,3,4}\left(
\int \frac{1}{E_{i}}d^{3}p_{i}\int_{\Sigma _{i}}p_{i\mu }d\sigma _{i}^{\mu
}\right)
\frac{f_{01}f_{02}f_{03}f_{04}}{\gamma _{1}\gamma _{2}\gamma
_{3}\gamma _{4}}\frac{g^{4}}{T^{4}}\left\langle \sigma _{1}\sigma _{2}\sigma
_{3}\sigma _{4}\right\rangle _{c}.
\end{align}
where the correlators of the sigma field can be derived from the probability distribution function with cubic
and quartic terms $P[\sigma ]=\exp \left\{ -\Omega \left[ \sigma \right] /T\right\}
   = \exp \left\{-\int d^{3}x\left[ \frac{1}{2}\left( \nabla
\sigma \right) ^{2}+\frac{1}{2}m_{\sigma }^{2}\sigma ^{2}+\frac{\lambda _{3}%
}{3}\sigma ^{3}+\frac{\lambda _{4}}{4}\sigma ^{4}\right]/T\right\}$~\cite{Stephanov:2008qz,Stephanov:2011pb}.
With such equilibrium distribution $P[\sigma]$, the deduced $\left\langle \left( \delta N\right) ^{2}\right\rangle _{c}$,
$\left\langle \left( \delta N\right) ^{3}\right\rangle _{c}$ and $\left\langle \left( \delta N\right) ^{4}\right\rangle _{c}$
belong to the category of static critical fluctuations.
If replacing the related integrations on the freeze-out surface by integrations over
the whole position space, the standard formula for a static and infinite medium
given by Stephanov in 2009~\cite{Stephanov:2008qz} can be reproduced~\cite{Jiang}.

To obtain the needed freezeout surface $\Sigma$, we implement the viscous hydrodynamic code
VISH2+1\cite{Song:2007fn} and extend its 2+1-d freezeout surface to the longitudinal direction
with the momentum and space rapidity correlations. For simplicity, we neglect succeeding
hadronic scatterings and resonance decays below $T_c$. We assume the critical and
noncritical fluctuations are independent, and
use the Poission and Binomial distributions as the non-critical fluctuations baselines.
Correspondingly, the total cumulants are expressed as: $C_n=(C_n)^{non-critical} + (C_n)^{critical}$, $n = 2, 3, 4$ 
(with $(C_n)^{critical}=\left\langle \left( \delta N\right) ^{n}\right\rangle _{c}$).
To roughly fit the trends of $C_2$, $C_3$ and $C_4$ of net protons,
we tune the couplings $g$, $\tilde{\lambda}_{3}$ $\left( \lambda _{3}=\tilde{\lambda}_{3}T\left( T\xi \right)
^{-3/2}\right)$ and $\tilde{\lambda}_{4}$ $\left( \lambda _{4}=\tilde{\lambda}
_{4}\left( T\xi \right) ^{-1}\right) $, and the correlation length $\xi$
within the allowed parameter ranges for each collision energy (please refer to~\cite{Jiang} for details).

\begin{figure*}[tbp]
\center
\begin{center}
{\normalsize \textbf{Au+Au 0-5\%, \ Thermal+Critical fluctuations (Poisson
baselines )} }
\end{center}
\par
%\vspace{-0.5cm}
\includegraphics[width=2.9 in]{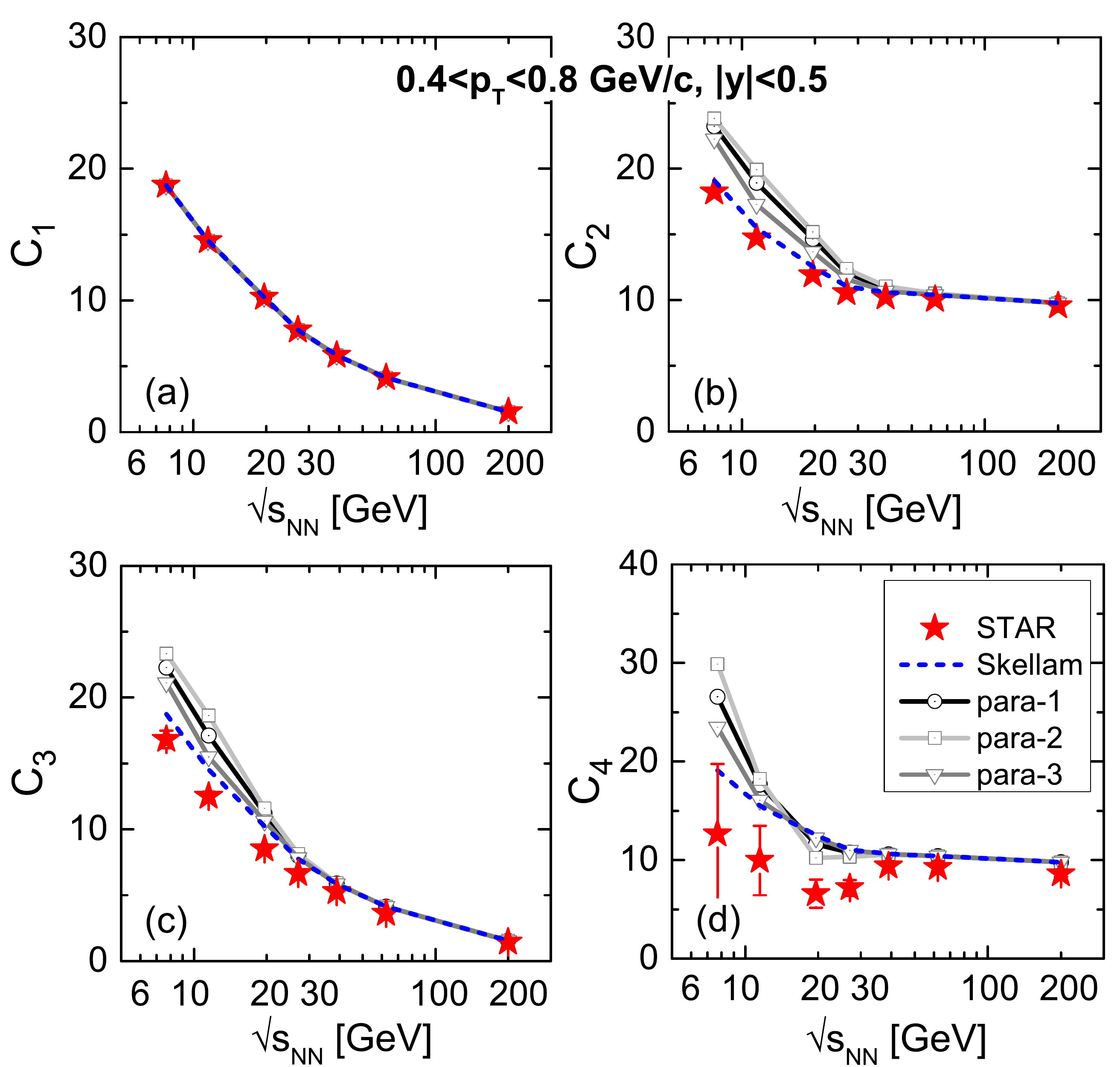} %
\includegraphics[width=2.9 in]{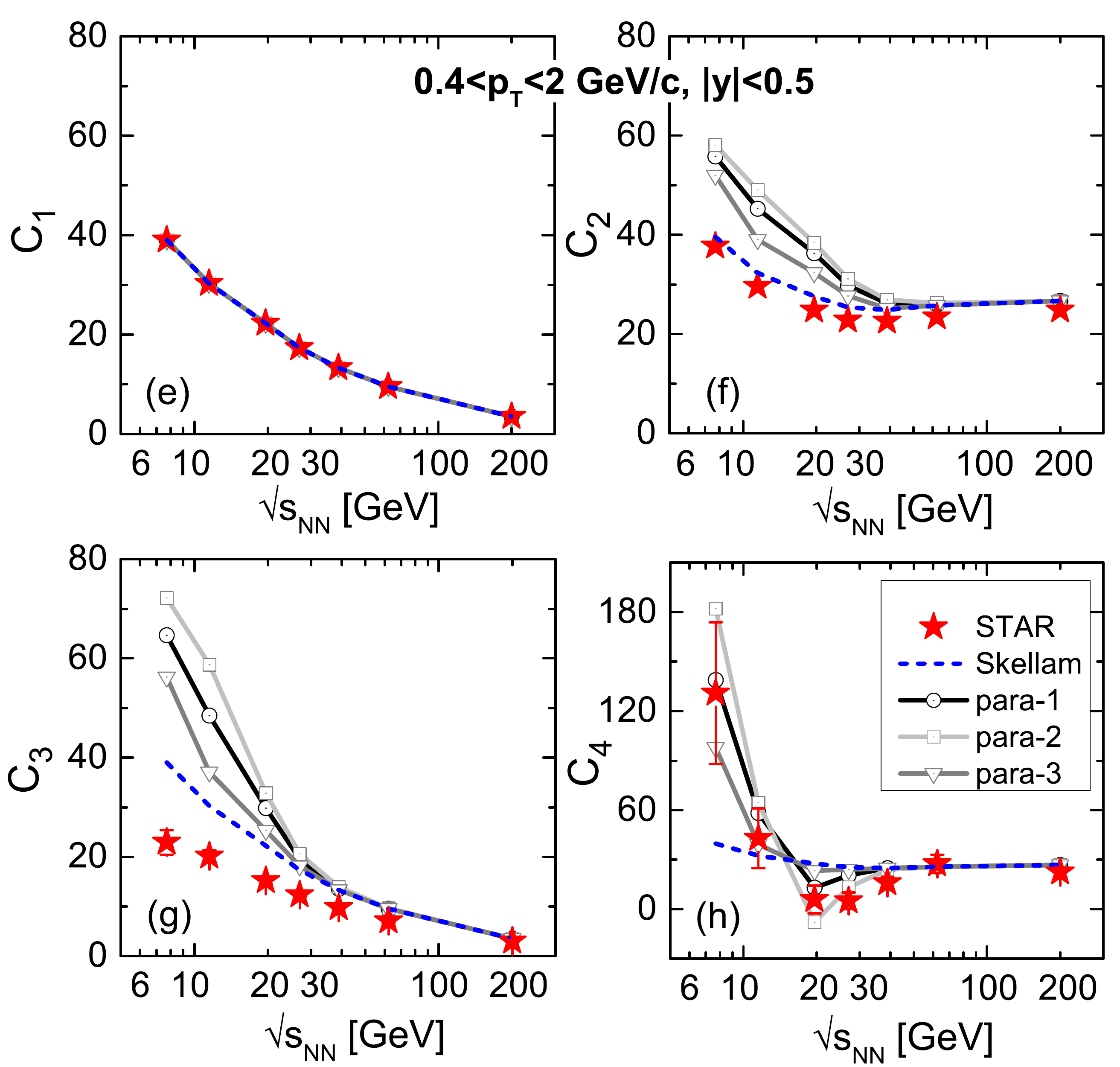}
\caption{Energy dependence of cumulants $C_1-C_4$  for net protons in 0-5\%  Au+Au collisions within $0.4 <p_T< 0.8\ \mathrm{GeV}$ (left panels) and within $0.4 <p_T< 2\ \mathrm{GeV}$ (right panels).  The red stars are the STAR preliminary data~\cite{Adamczyk:2013dal,Luo:2015ewa}, dashed blue lines are the Poisson expectations, and black and grey curves with symbols are the results from our model calculations with Poisson baselines.}
\label{cumulant4-ske-0005}
\vspace{0.5cm}
\begin{center}
{\normalsize \textbf{Au+Au 0-5\%, \ Thermal+Critical fluctuations (Binomial
baselines )} }
\end{center}
\center
\includegraphics[width=2.9 in]{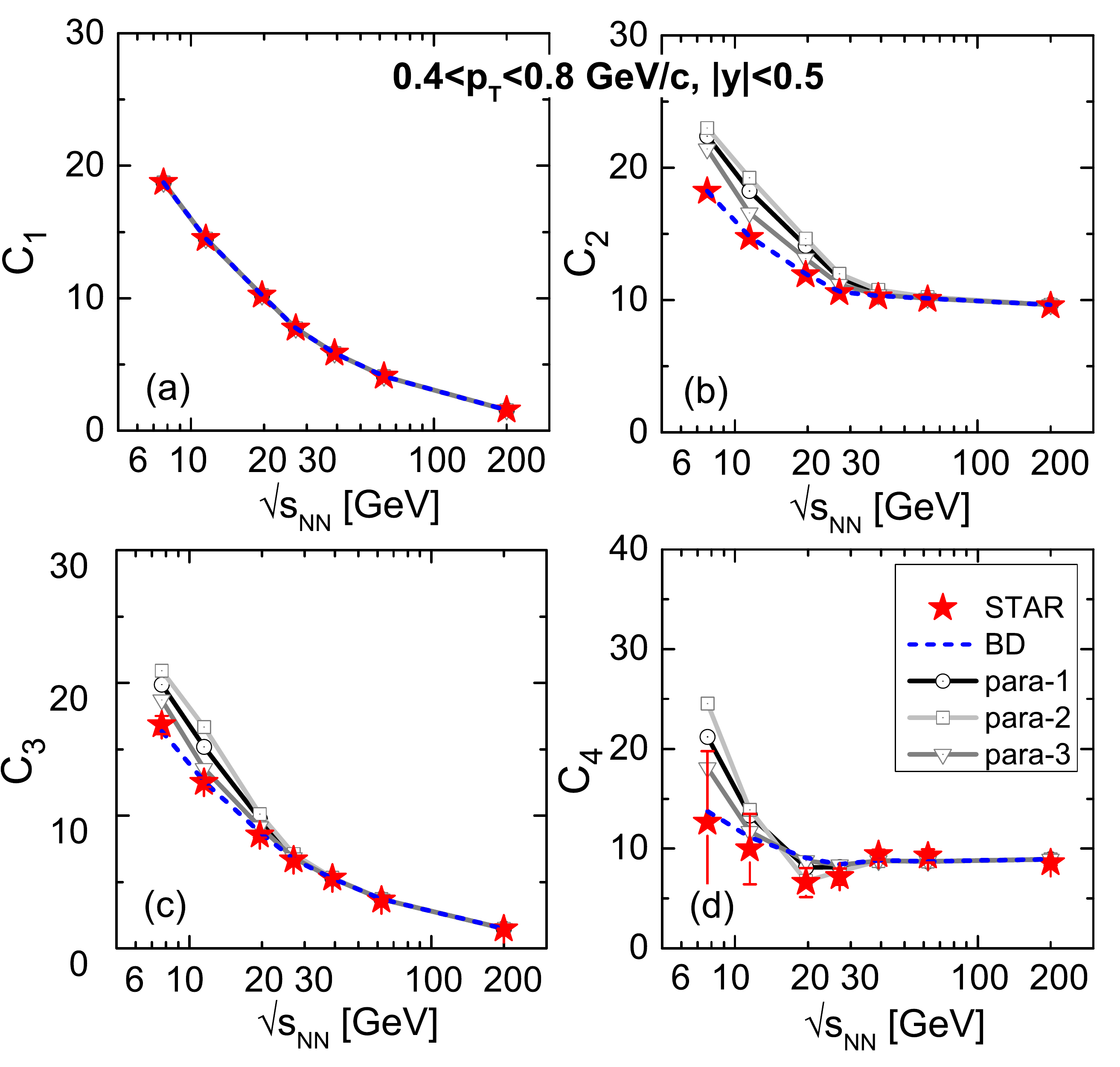}
\includegraphics[width=2.9 in]{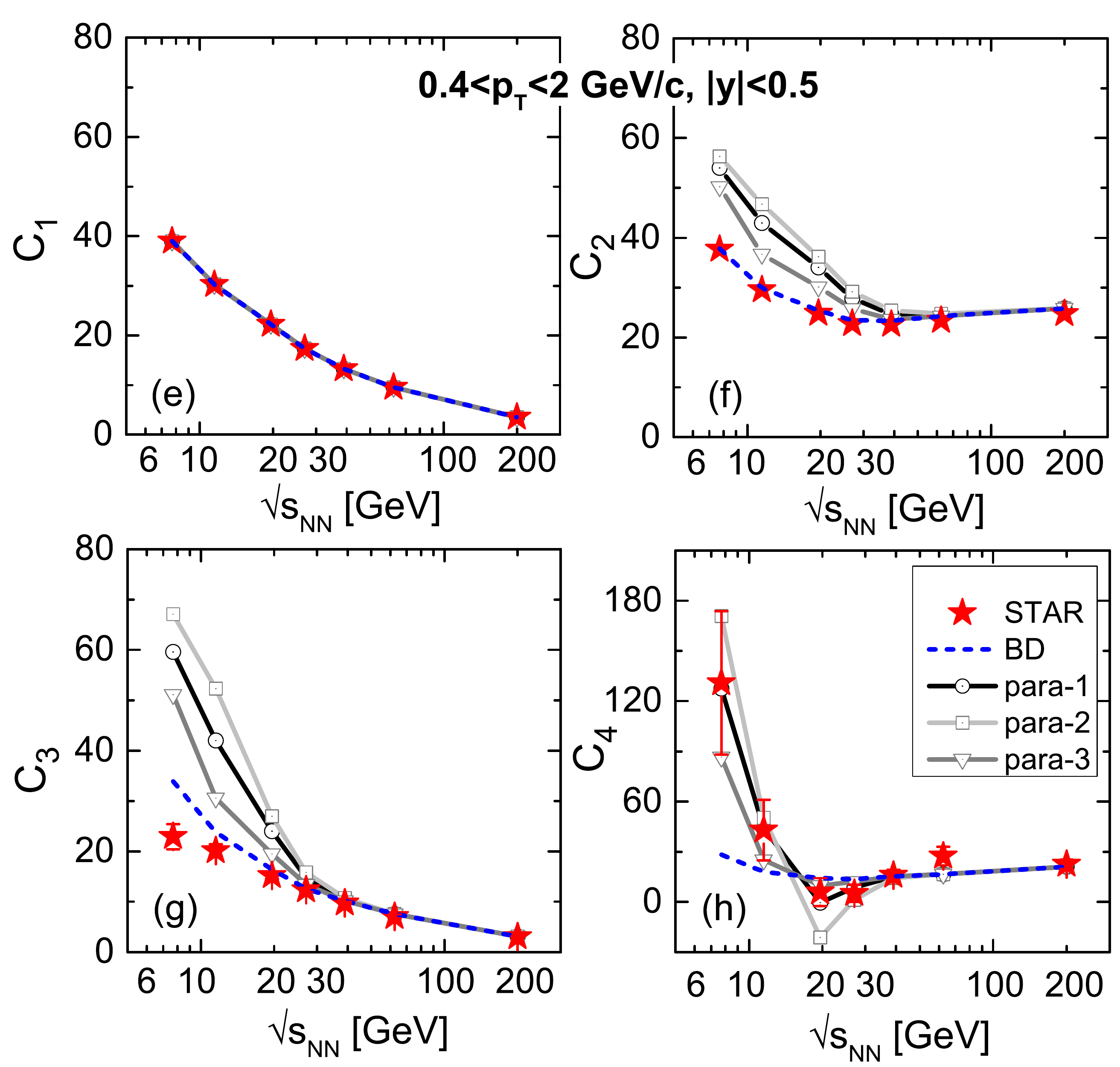}
\caption{Similar to Fig.1, but with Binomial baselines.}
\label{cumulant4-ske-0005}
\end{figure*}

\begin{figure*}[tbp]
\center
\includegraphics[width=2.9 in]{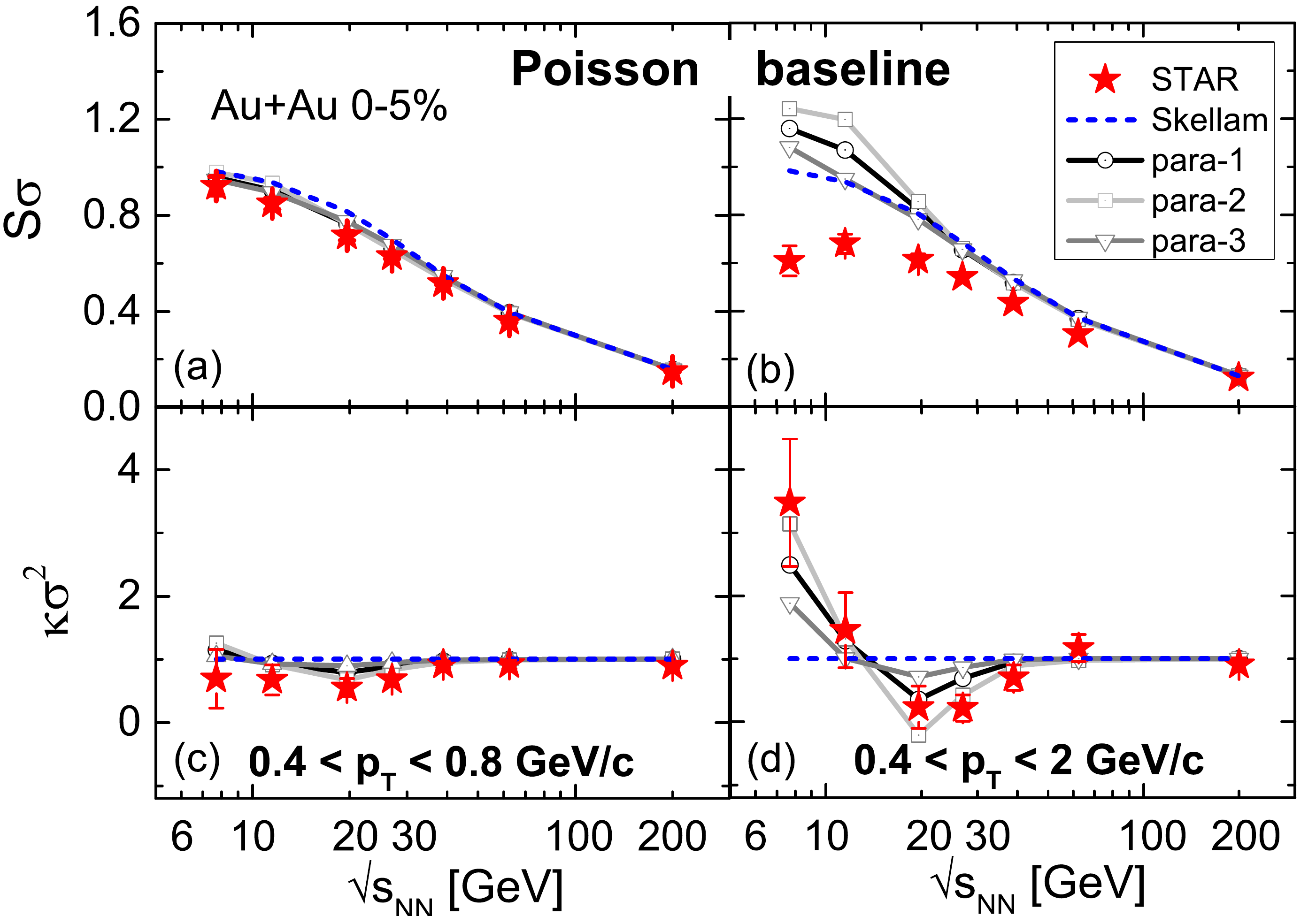}
\includegraphics[width=2.9 in]{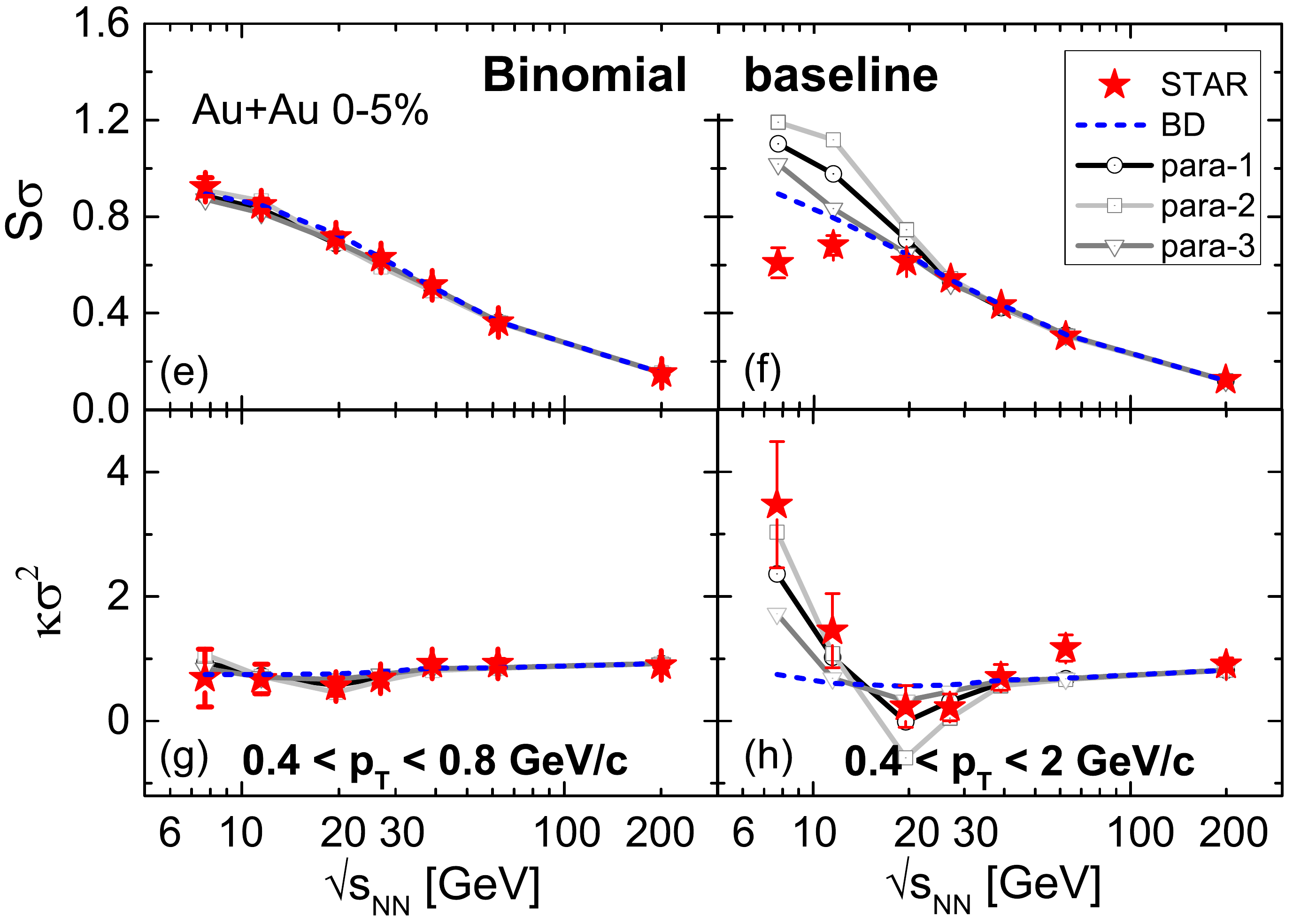}
\caption{Energy dependence of cumulant ratios, $S \sigma$ and $\kappa \sigma^2$, for net protons in  0-5\%
Au+Au collisions.}
\label{c-ratio-1}
\end{figure*}

\section{Numerical results}

Fig.~1 and Fig.~2 show the energy dependence of cumulants $C_{1}-C_{4}$ for
net protons in the most central Au+Au collisions with either Poisson or Binomial baselines.
After tuning $g$, $\xi$, $\tilde{\lambda}_{3}$
and $\tilde{\lambda}_{4}$ within the allowed parameter ranges, we roughly
describe the decreasing trend of $C_2$ and $C_3$ and the non-monotonic
behavior of $C_4$ with the increase of collision energy. However, $C_{2}$
and $C_{3}$ from our model calculations are always above the
Poisson/Binomial baselines due to the positive contributions from the critical fluctuations.
For the Binomial baselines, our model calculations can nicely fit the energy dependent $C_{4}$
within two different $p_{T}$ ranges. However, if using the Poisson baselines, our calculations
can not simultaneously describe the $C_{4}$ data at lower collision energies.
For Au+Au collisions below 11.5 GeV, the measured $C_{4}$ are higher than
the Poisson expectation values for $0.4<p_{T}<2 \ \mathrm{GeV}$, but lower than
the Poisson expectation values for $0.4<p_{T}<0.8 \ \mathrm{GeV}$. For Eqs.(2-4),
the change of the $p_{T}$ ranges only affects the magnitude of the $C_{n}^{critial}$ from the
critical fluctuations, rather than their signs, which thus can not explain the $C_{4}$ data at
lower collision energies with the Poisson baseline.

Fig.~1 and Fig.~2 also show that, with the maximum $p_{T}$ increased from 0.8 GeV to 2 GeV,
the higher cumulants from both experiment and model dramatically increase, showing large deviations from
the Poisson/Binomial baselines. In fact,
the $n_{th}$ order critical fluctuations from Eqs.(2-4) are closely related to the $n_{th}$ power of the
total net-proton numbers within the defined $p_T$ and rapidity range. With
the maximum $p_T$ increased from $0.8 \ \mathrm{GeV}$ to $2 \ \mathrm{GeV}$,
the averaged numbers of the net protons almost increase by a
factor of two, leading to large increase of higher cumulants in our calculations.

Fig.~3 shows the energy dependence of cumulant ratios $S\sigma
=C_{3}/C_{2}$ and $\kappa \sigma ^{2}=C_{4}/C_{2}$ in 0-5\%
Au+Au collisions with Poisson/Binomial baselines. Although our model calculations
over-predict $C_{2}$ and $C_{3}$, the cumulants ratios $S\sigma $ and $%
\kappa \sigma ^{2}$ show better agreement with the experimental measurements
in the most central collisions, except for $S\sigma $ with
$0.4<p_{T}<2 \ \mathrm{GeV}$.
Ref~\cite{Jiang} also showed that the
critical fluctuations dramatically decreased from most central to semi-peripheral Au+Au collisions.
For 30-40\% centrality, our model calculations (in \cite{Jiang}) can nicely fit $S\sigma$ and $\kappa \sigma ^{2}$
with the Binomial baselines, but fail
to fit $\kappa \sigma ^{2}$ with the Poisson
baselines since the Poisson expectations largely deviate from the experimental data there.

\section{summary}

Based on Ref.~\cite{Jiang}, this proceeding briefly introduced the freezeout scheme near the QCD critical point and outlined
the formulism to calculate the correlated fluctuations of net protons on the hydrodynamic freeze-out surface
with the presence of an external order parameter field.  Our model calculations could roughly
describe the decrease trend of $C_2$ and $C_3$ and the non-monotonic behavior of $C_4$ and $\kappa \sigma^2$ through tuning the related
parameters, but always over-predict
the values of $C_{2}$ and $C_{3}$ for both Poisson and  Binomial baselines due to the positive contributions from
the static critical fluctuations. To solve this sign problem of $C_{2}^{critcal}$ and $C_{3}^{critcal}$,
the dynamical evolution of the sigma field and more realistic thermal fluctuation
baselines should be investigated in the near future.\\[-0.08in]

$Acknowledgement:$ This work is supported by the NSFC and the MOST under grant
Nos. 11435001 and 2015CB856900.

\vspace{-2mm}

\end{document}